\documentclass[hidelinks,11pt]{article}

\newcommand{\bol}{\boldsymbol}

\newcommand{\ney}{\boldsymbol{y}}                          
\newcommand{\nex}{\boldsymbol{x}}

\newcommand{\ner}{\boldsymbol{r}}

\newcommand{\de}{\,\mathrm{d}}                               
\newcommand{\e}{\operatorname{e}}                               
                     
\newcommand{\inc}{\mathrm{inc}}

\newcommand{\andtext}{\quad\mbox{and}\quad}

\newcommand{\p}{\partial}

\newcommand{\real}{\mathrm{Re}\,}    
                                   
\newcommand{\imag}{\mathrm{Im}\,}

\newcommand{\lf}{\left}
\newcommand{\rg}{\right}

\newcommand{\R}{\mathbb{R}}       
\newcommand{\C}{\mathbb{C}}

\usepackage{multirow}
\usepackage{amsmath}
\usepackage{amsfonts}
\usepackage{amsmath}
\usepackage{amssymb}  
\usepackage{graphicx}
\usepackage{caption}
\usepackage{mathrsfs}
\usepackage{upgreek}
\usepackage{amsthm}
\usepackage{subfig}
\usepackage{booktabs}
\usepackage{authblk}
\usepackage{bm}

\newtheorem{theorem}{Theorem}[section]

\newtheorem{remark}[theorem]{Remark}

\title{Convolution quadrature methods for time-domain  scattering from unbounded penetrable interfaces}
\author[1]{Ignacio Labarca}
\affil[1]{\small{Institute for Mathematical and Computational Engineering, School of Engineering and Faculty of Mathematics, Pontificia Universidad Cat\'olica de Chile, Santiago, Chile}}
\author[2]{Luiz Faria}
\affil[2]{\small{Department of Mathematics, Massachusetts Institute of Technology, Cambridge, MA, USA}}
\author[1.2]{Carlos P\'erez-Arancibia\thanks{cperez@mat.uc.cl, cperezar@mit.edu}}
\date{\today}
\date{\today}

\begin{document}
\maketitle


\begin{abstract}
This paper presents a class of boundary integral equation methods for the numerical solution of  acoustic and electromagnetic  time-domain scattering problems in the presence of unbounded penetrable interfaces in two-spatial dimensions. The proposed methodology relies on Convolution Quadrature (CQ) methods in conjunction with the recently introduced Windowed Green Function (WGF) method. As in standard time-domain scattering from bounded obstacles, a CQ method of the user's choice is utilized to transform the problem into a finite number of (complex) frequency-domain problems posed on the domains involving penetrable unbounded interfaces. Each one of the  frequency-domain transmission problems is then formulated as a second-kind integral equation that is effectively reduced to a bounded interface by means of the WGF method---which introduces errors that decrease super-algebraically fast as the window size increases. The resulting windowed integral equations can then be solved by means of any (accelerated or unaccelerated) off-the-shelf Helmholtz boundary integral equation solver capable of handling complex wavenumbers with a large imaginary part. A high-order Nystr\"om method based on Alpert quadrature rules is utilized here. A variety of numerical examples including wave propagation in open waveguides as well as scattering from multiply layered media, demonstrate the capabilities of the proposed approach.  
\end{abstract}

\maketitle 

\section{Introduction} 
Wave propagation problems involving unbounded material interfaces play a fundamental role in numerous relevant electromagnetic and acoustic engineering applications such as waveguides, solar cells, on-chip antennas, and more recently, inverse metasurface design, to mention a few. Typically, frequency- and time-domain simulations in this context are performed by means of volume discretization techniques such as finite difference (FDTD~\cite{oskooi2010meep} or TDFD~\cite{taflove2005computational} methods) and finite element~\cite{jin2015finite} methods where perfectly matched layers (PMLs)~\cite{berenger1994perfectly} or other kinds of absorbing/transparent boundary conditions are used to reformulate the problem in a \emph{bounded} domain free of \emph{unbounded} material interfaces. 

Time-domain boundary integral equations for obstacle scattering problems, on the other hand, have been extensively and intensively studied over the last two decades~\cite{dominguez2017recent}. Convolution quadrature (CQ) methods~\cite{lubich1988convolution,lubich1988convolution_II}, in particular, have effectively enabled the use of (complex) frequency-domain boundary integral equation (BIE) solvers to tackle a variety of wave propagation problems, by providing a stable procedure to discretize the associated convolution equations for the unknown time evolution of the relevant surface densities; see~\cite{sayas2016retarded} for the mathematical foundations of the method, and \cite{banjai2012wave,hassell2016convolution} for details on the algorithmic implementation. In the case of the scalar wave equation with piecewise constant wavespeed, to which this paper is devoted to, approximate traces at discrete times are produced all at once from a finite sequence of independent Helmholtz problems that can be solved in parallel by means of BIE methods. Although this CQ-BIE approach has proven to be competitive to volume discretization methods in the context of obstacle scattering problems~\cite{banjai2014fast,Banjai:2009in,schadle2006fast}, its extension to problems involving unbounded material interfaces is severely hindered by the fact that standard BIE formulations require the knowledge of problem-specific Green functions to deal with the unboundedness of the material interfaces. These Green functions, however, are often unavailable (in terms of tractable mathematical expressions) or are given in terms of computationally expensive Sommerfeld integrals\footnote{The numerical evaluation of Sommerfeld integrals has been referred to in the literature as ``a standard nightmare  for many electromagnetic engineers"~\cite{jimenez1996sommerfeld}}~\cite{michalski2016efficient,perez2014high,perez2017windowed}.

Recent advances on BIE methods for time-harmonic problems of scattering from unbounded material interfaces have led to the development of highly efficient solvers that completely bypass the use of problem-specific Green functions~\cite{bruno2017guide,bruno2016windowed,bruno2017windowed,lu2018perfectly,perez2017windowed,zhang2011novel}. The windowed Green function (WGF) method, in particular,  has successfully been used in layered media~\cite{bruno2016windowed,bruno2017windowed,perez2017windowed},  dielectric waveguides~\cite{bruno2017guide} and all-dielectric metasurfaces~\cite{pestourie2018inverse} simulations in the frequency domain. The method
relies on a certain ``second-kind" BIE---given in terms of free-space Helmholtz Green functions---posed on all the (bounded and unbounded) interfaces. A highly accurate approximate solution to this BIE is then obtained by solving a modified \emph{windowed BIE} posed on the relevant \emph{bounded} portions of material interfaces. The windowed BIE is directly obtained from the original BIE by simply multiplying the integral kernels by a smooth \emph{window function}. The resulting windowed BIE is (provable) of the second-kind, it is given in terms of the four standard BIE operators of Calder\'on calculus, and it thus can be solved by means of any (accelerated or unaccelerated) off-the-shelf  Nystr\"om~\cite{bruno2009electromagnetic,bruno2001fast,colton2012inverse} or boundary element method (BEM)~\cite{sauter2010boundary} solver.

This paper presents a combined CQ-WGF procedure for problems of time-domain scattering from unbounded material interfaces  ruled by the scalar wave equation in two-spatial dimensions. Our goal is to show that the straightforward combination of the two methods suffices to extend the reach of efficient BIE solvers to the large class of relevant engineering problems involving unbounded penetrable interfaces. The proposed procedure is simple. At first, a CQ method is utilized to turn the scalar wave equation into a finite sequence of frequency-domain transmission problems in the domains containing the unbounded penetrable interfaces.  Each one of the required frequency-domain problems is then  formulated as ``second-kind" indirect BIE which is approximated---with errors that decay super-algebraically fast as the window size increases---by means of the WGF method, which is applicable for the complex wavenumbers produced by the CQ method~\cite{bruno2016windowed,perez2017windowed}. For the sake of definiteness we consider here the FFT-accelerated CQ method put forth in~\cite{Banjai:2009in} and the high-order BIE Nystr\"om method described in~\cite{Hao:2013do} which based on Alpert quadrature rules~\cite{alpert1999hybrid}.  The capabilities of the proposed procedure are demonstrated by a variety of numerical examples including wave propagation in open waveguides and waveguide branches, as well as scattering from multiply layered media.

The structure of this paper is as follows: Section~\ref{sec:set-up} sets forth the model problem used throughout this paper to described the proposed combined CQ-WGF methodology. Section~\ref{sec:CQ} details the original CQ method, based on linear multi-step methods, applied to our model problem. Next, in Section~\ref{sec:WGFM} the BIE formulation of the frequency-domain problems and the WGF method are presented. Section~\ref{sec:num}, finally, contains the solver validation and the numerical results corresponding to the examples considered.

\begin{figure}[h!]
\centering	
\includegraphics[scale=1]{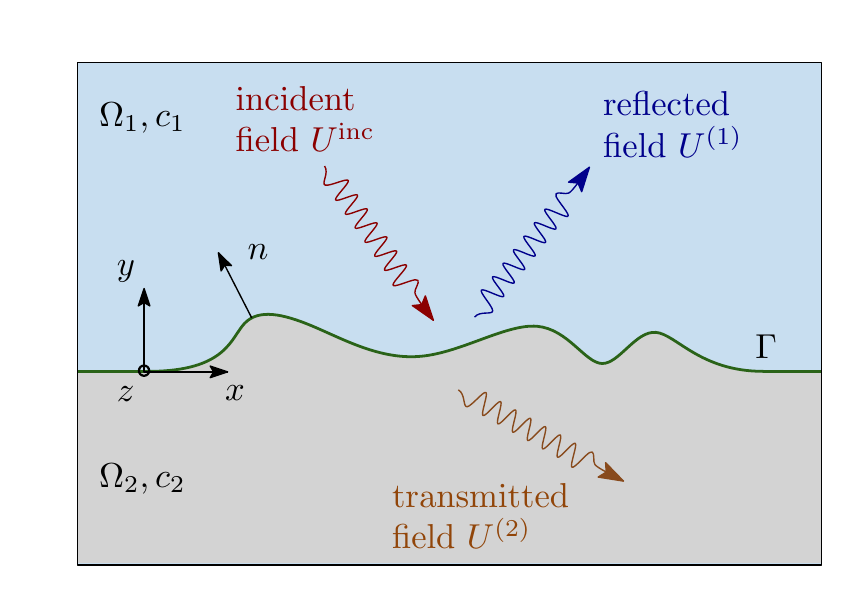}
\caption{Geometry of the model problem: A time-dependent incident field ($U^{\inc}$) impinges on a locally perturbed half-plane producing a reflected ($U^{(1)}$) and a transmitted ($U^{(2)}$) field. }\label{fig:dom}
\end{figure}

\section{Model problem\label{sec:set-up}}
This section is devoted to set up the model problem used for the presentation of the proposed technique.  Without loss of generality we focus here on the electromagnetic scattering problem; an analogous acoustic problem can also be formulated. Consider then the locally perturbed dielectric half-plane depicted in Figure~\ref{fig:dom}. The upper and lower media are denoted by $\Omega_1$ and  $\Omega_2$, within which the wavespeed equals $c_1=(\mu_1\epsilon_1)^{-1/2}>0$ and $c_2=(\mu_2\epsilon_2)^{-1}>0$, respectively, with $\mu_j$ and $\epsilon_j$  denoting the magnetic permeability and the electric permittivity of the dielectric medium $\Omega_j$, $j=1,2$. The common unbounded interface between the two media is denoted by $\Gamma$ and is assumed to be a piecewise smooth curve. We then consider a TE- or TM-polarized incident electromagnetic field $U^\inc$ that impinges on the  interface  $\Gamma$ producing a reflected and a transmitted field, as is depicted in Figure~\ref{fig:dom}.  The scalar field~$U^\inc$---which satisfies the wave equation $\p_t^2U^\inc(\nex,t)-c_1^2\Delta U^\inc(\nex,t)=0$ in all of $\R^2\times\R_+$---denote $z$-component of either the incident electric field in TE-polarization or the incident magnetic field in TM-polarization. It is assumed that the incident field arrives at $\Gamma$ at a time $t_0>0$ so that both the reflected field in $\Omega_1$ and the transmitted field in $\Omega_2$ equal zero at~$t=0$. 

Expressing the $z$-component of the resulting total electromagnetic field as 
$$
U =\lf\{\begin{array}{ccc}U^{(1)}+U^\inc&\mbox{in}&\Omega_1,\\
U^{(2)}&\mbox{in}&\Omega_2,\end{array}\right. 
$$
we then obtain that the reflected  and transmitted fields---which are denoted by $U^{(1)}$ and $U^{(2)}$, respectively---satisfy
\begin{subequations}\begin{eqnarray}
 \p^2_tU^{(j)}(\nex,t)-c_j^2\Delta U^{(j)}(\nex,t)&=&0,\quad   (\nex,t)\in\Omega_j\times \R_+,\quad j=1,2,\label{eq:wave}\\
 U^{(2)}(\nex,t) -U^{(1)}(\nex,t)&=&U^\inc(\nex,t),\quad(\nex,t)\in\Gamma\times\R_+,\label{eq:dir}\\
 \nu_2 \p_n U^{(2)}(\nex,t)-\nu_1\p_n U^{(1)}(\nex,t) &=& \nu_1\p_nU^\inc(\nex,t),\quad(\nex,t)\in\Gamma\times\R_+,\label{eq:neu}\\
  U^{(j)}(\nex,0) =\p_tU^{(j)}(\nex,0)&=&0,\quad\nex\in\Omega_j,\quad j=1,2,\label{eq:init}
\end{eqnarray}\label{eq:wave-eqn}\end{subequations}
where $\nu_j=\mu^{-1}_j$ in TE-polarization and $\nu_j = \epsilon^{-1}_j$ in TM-polarization. 

In the following section we present the multi-step time semi-discretization of the transmission problem~\eqref{eq:wave-eqn}  using  the classical CQ method introduced by Lubich in~\cite{lubich1988convolution}.

\section{Convolution quadrature methods\label{sec:CQ}}
Following  the presentation of the convolution quadrature in~\cite{betcke2017overresolving}, we begin by turning the second-order  transmission problem~\eqref{eq:wave-eqn} into a first-order system. We thus introduce the vector valued functions $\bold V^{(j)}(\nex,t ) = \left[U^{(j)}(\nex,t),c_j^{-1}\p_tU^{(j)}(\nex,t)\right]^T$, $j=1,2,$ which allow~\eqref{eq:wave-eqn} to be expressed as
\begin{subequations}\begin{eqnarray}
 c_j^{-1}\p_t \bold V^{(j)}(\nex,t)&=&\mathcal L\bold V^{(j)}(\nex,t),\quad   (\nex,t)\in\Omega_j\times \R_+,\quad j=1,2,\label{eq:1_ODE}\\
 \mathcal B_2\bold V^{(2)}(\nex,t)&=& \mathcal B_1\bold V^{(1)}(\nex,t)+\bold F(\nex,t),\quad   (\nex,t)\in\Gamma\times \R_+, \\
  \bold V^{(j)}(\nex,0) &=&\bold 0, \quad\nex\in\Omega_j,\quad j=1,2,
\end{eqnarray}\label{eq:1st_ODE_mod}\end{subequations}
where $\mathcal L=\begin{bmatrix}0&I\\\Delta&0\end{bmatrix},$ $\mathcal B_j=\begin{bmatrix}\gamma_D&0\\0&\nu_j\gamma_N\end{bmatrix}$ and $\bold F(\nex,t)=\begin{bmatrix}U^\inc(\nex,t)\\\nu_1\p_nU^\inc(\nex,t)\end{bmatrix}$. (Note that $\bold F(\nex,\cdot)$ is a causal function.) The symbols $\gamma_D$ and $\gamma_N$ in the definition of the operators $\mathcal B_j$, $j=1,2$, denote the Dirichlet and Neumann traces on~$\Gamma$, respectively.

The system~\eqref{eq:1st_ODE_mod} is subsequently semi-discretized in time using a general linear multi-step method. Letting $\Delta t>0$ denote the prescribed time step and $\alpha_\ell,\beta_\ell\in\R$, $\ell=0,\ldots,k,$ denote the coefficients of the multi-step method, we obtain that equations~\eqref{eq:1_ODE} for $j=1,2$, become the following difference equations
 \begin{equation}\begin{split}
 \frac{1}{c_j\Delta t}\sum_{\ell=0}^k\alpha_{\ell}\bold V^{(j)}_{n+\ell-k}(\nex) = \sum_{\ell=0}^{k}\beta_{j}\mathcal L\bold  V^{(j)}_{n+\ell-k}(\nex),
\end{split}\label{eq:LMS}\end{equation}  where we have introduced the sequences of vector valued functions $\lf\{\bold V_n^{(j)}(\cdot)\rg\}_{n=-\infty}^\infty$, $j=1,2$, which correspond to the approximation $\bold V^{(j)}(\cdot,t_n)\approx \bold V^{(j)}_{n}(\cdot)$ at the discrete times $t_n=n\Delta t$  for $n\geq 0$, and are defined as $\bold V^{(j)}_{n}(\cdot)=\bold 0$ for~$n<0$.

As it turns out, the difference equations~\eqref{eq:LMS}  can be solved by means of the $\zeta$-transform~\cite{hassell2016convolution}. Indeed, applying $\zeta$-transform to both sides of~\eqref{eq:LMS} we get
$$
\frac{1}{c_j\Delta t}\sum_{n=0}^{\infty}\lf(\sum_{\ell=0}^k\alpha_{\ell}\bold V^{(j)}_{n+\ell-k}(\nex)\rg)\zeta^n = \sum_{n=0}^\infty \lf(\sum_{\ell=0}^{k}\beta_{\ell}\mathcal L\bold  V^{(j)}_{n+\ell-k}(\nex)\rg)\zeta^n,
$$
for $\zeta\in B\subset\C$ with $B$ denoting the region of convergence of the power series. From the convolution property of the $\zeta$-transform it  follows that the functions $\bold v^{(j)}(\cdot,\zeta)$, $j=1,2$---which correspond to  the $\zeta$-transform of the sequences $\lf\{\bold V^{(j)}(\cdot)\rg\}_{n=-\infty}^{\infty}$, $j=1,2$---satisfy 
\begin{subequations}\begin{eqnarray}
\lf(\frac{\gamma(z)}{c_j\Delta t}\rg)\bold v^{(j)}(\nex,\zeta) &=&\mathcal L\bold v^{(j)}(\nex,\zeta),\qquad(\nex,\zeta)\in\Omega_j\times B,\ j=1,2,\\
 \mathcal B_2\bold v^{(2)}(\nex,\zeta)&=&\mathcal B_1\bold v^{(1)} (\nex,\zeta)+\bold f(\nex,\zeta),\quad   (\nex,\zeta)\in\Gamma\times B, 
\end{eqnarray}\label{eq:discrete_equation}\end{subequations}
where   $\gamma(\zeta)=\lf(\sum_{\ell=0}^k \alpha_\ell \zeta^{k-\ell}\rg)/\lf(\sum_{\ell=0}^k \beta_\ell \zeta^{k-\ell}\rg)$,  
\begin{equation}
 \bold v^{(j)}(\nex,\zeta) = \sum_{n=0}^{\infty}\bold V^{(j)}_n(\nex)\zeta^{n} \andtext \bold f(\nex,\zeta) = \sum_{n=0}^{\infty}\bold F(\nex,n\Delta t)\zeta^{n}.\label{eq:z_trans}
\end{equation}
 Letting $\bold v^{(j)}=[u^{(j)},v^{(j)}]^T$ and $\bold f=[f,g]^T$ it readily follows that the scalar fields $u^{(j)}:\Omega_j\times B\to\C$, $j=1,2$, satisfy the Helmholtz transmission problem:
\begin{subequations}\begin{eqnarray}
\Delta u^{(j)}(\nex,\zeta) + k_j^2(\zeta) u^{(j)}(\nex,\zeta) &=&0, \ \quad\qquad  (\nex,\zeta)\in \Omega_j\times B,\quad j=1,2,\label{eq:helmholtz}\\
u^{(2)}(\nex,\zeta) -u^{(1)}(\nex,\zeta) &=& f(\nex,\zeta), \quad  (\nex,\zeta)\in \Gamma\times B,\label{eq:trans_cond_1}\\
\nu_2\p_nu^{(2)}(\nex,\zeta)-\nu_1\p_nu^{(1)}(\nex,\zeta)  &=&   g(\nex,\zeta), \quad (\nex,\zeta)\in \Gamma\times B,\label{eq:trans_cond_2}
\end{eqnarray}\label{eq:transmission}\end{subequations}
where the (complex) wavenumbers $k_j(\zeta)$, $j=1,2,$ are given by 
\begin{equation}
k_j(\zeta) :=  \frac{i\gamma(\zeta)}{c_j\Delta t}.\label{eq:comple_wn}
\end{equation}
The transmission problem~\eqref{eq:transmission} needs to be complemented with suitable radiation conditions for both  fields $u^{(1)}$ and $u^{(2)}$ that ensure that they correspond to waves that propagate away from the interface~$\Gamma$. We refer the reader to~\cite{KRISTENSSON:1980vs} for a rigorous discussion on suitable radiation conditions that lead to results on the existence and uniqueness  of solutions of~\eqref{eq:transmission} for physically meaningful wavenumbers, i.e., $k_j(\zeta)\in\C$ such that  $\real k_j(\zeta)>0$ and $\imag k_j(\zeta)\geq 0$. As it turns out, only transmission problems~\eqref{eq:transmission} with wavenumbers satisfying the latter conditions are needed to be solved for the implementation of the proposed convolution quadrature method (see Remark~\ref{rem:pos_wn} below). In Section~\ref{sec:WGFM} below, we present an efficient and high-order boundary integral method for the fast and accurate  solution of the transmission problems~\eqref{eq:transmission}.

Assuming that the (complex) frequency-domain solutions~$u^{(j)}$, $j=1,2,$ have been obtained by solving~\eqref{eq:transmission}, the approximations $U^{(j)}_n(\nex)\approx U^{(j)}(\nex, t_n)$ for the reflected ($j=1$) and transmitted ($j=2$) fields at the discrete times $t_n=n\Delta t$, $n=0,\ldots N,$ are  retrieved by taking inverse $\zeta$-transform of $u^{(j)}(\nex,\cdot)$, $j=1,2$. It thus follows directly from~\eqref{eq:z_trans} and the Cauchy residue theorem that $u^{(j)}_n(\nex)$ can be expressed as the complex contour integrals
\begin{equation}
U^{(j)}_n(\nex) :=\frac{1}{2\pi i}\oint_C\frac{u^{(j)}(\nex,\zeta)}{\zeta^{n+1}}\de \zeta,\quad n=0,\ldots, N,\quad  \nex\in\Omega_j\quad j=1,2,\label{eq:inv_Z}
\end{equation}
where the contour $C$ could be any simple closed curve contained in $B$.  Note that the validity of  formula~\eqref{eq:inv_Z} relies on the  analyticity of $u^{(j)}(\nex,\cdot)$ within the region enclosed by $C$ in complex $\zeta$-plane. Results on these regards for  the transmission problem~\eqref{eq:transmission} can be derived following the arguments presented in~\cite{cutzach1998existence}. 
We point out here that no scattering poles associated with~\eqref{eq:transmission} lie inside the contour for $C\subset\{z\in\C:\imag z\geq 0\}$. (An interesting discussion on analytic and numerical issues arising due to the existence of scattering poles near the contour $C$ utilized in the practical implementation of the convolution quadrature method can be found in reference~\cite{betcke2017overresolving}.)

In practice, the contour integrals~\eqref{eq:inv_Z} have to be computed numerically and any quadrature rule could be utilized in principle. As it is pointed out in reference~\cite{Banjai:2009in}, however, the use of the classical trapezoidal rule  leads to a significant reduction in the overall computational cost of the CQ method. To see this we proceed to select the contour $C$ as a circle of radius $\lambda>0$ which is discretized using the quadrature points $\zeta_m = \lambda \e^{2i\pi  m/(N+1)}$, $m=0,\ldots,  N$, that produce the following approximation of the integrals in~\eqref{eq:inv_Z}:
\begin{equation}
\widetilde U_n^{(j)}(\nex) := \frac{\lambda^{-n}}{N+1}\sum_{m=0}^Nu^{(j)}(\nex,\zeta_m)\zeta^{-{n}}_m, \quad n=0\ldots N,\quad j=1,2,\quad\nex\in\Omega_j.\label{eq:inv_Z_FFT}
\end{equation}
The advantages of the trapezoidal rule are twofold. On one hand the numerical errors in the approximations $U^{(j)}_n(\nex)\approx\widetilde U_n^{(j)}(\nex)$, $n=0,\ldots, N$, decay exponentially fast as $N$ increases (due to the analyticity and periodicity of the integrands in~\eqref{eq:inv_Z}), and, on the other hand, the sums in~\eqref{eq:inv_Z_FFT} for $n=0,\ldots,N,$ can be efficiently computed by means of the Fast Fourier Transform (FFT).

\begin{remark}\label{rem:pos_wn}The computational cost  associated to the evaluation of~\eqref{eq:inv_Z_FFT} can be further reduced by noticing that, using the complex conjugation, just half of the fields  $u^{(j)}$ corresponding to  solutions of~\eqref{eq:transmission} for wavenumbers $k_j(\zeta)$~\eqref{eq:comple_wn} in the first quadrant (i.e., $\real k_j(\zeta)>0$ and $\imag k_j(\zeta)\geq 0$) need to be computed~\cite{Banjai:2009in}.
\end{remark}

\section{Windowed Green function method}\label{sec:WGFM}
In this section we present the WGF method  for the solution of the two-layer transmission problem~\eqref{eq:transmission}. As is shown in~\cite{bruno2017guide,bruno2017windowed,perez2017windowed} and in the numerical examples presented in Section~\ref{sec:num}, the proposed WGF method approach can be easily extended to tackle more general configurations involving unbounded material interfaces, such as multiply layered media and waveguide branches. 

We first introduce the single- and double-layer potentials which are defined as
\begin{equation}
\lf(\mathcal S_{\zeta,j}\varphi\rg) (\ner) := \int_{\Gamma} G_{\zeta,j}(\ner,\ney)\varphi(\ney)\de s(\ney)\mbox{ and } \lf(\mathcal D_{\zeta,j}\varphi\rg) (\ner) := \int_{\Gamma} \frac{\p G_{\zeta,j}(\ner,\ney)}{\p n(\ney)}\varphi(\ney)\de s(\ney),\  \ner\in\R^2\setminus\Gamma,\label{eq:lay_pot}
\end{equation}
respectively, where 
$ G_{\zeta,j}(\nex,\ney):=\frac{i}{4}H_0^{(1)}(k_j(\zeta)|\nex-\ney|)\label{eq:GF}$ is the free space Green function for the Helmholtz equation with  wavenumbers $k_j=k_j(\zeta)$ defined in~\eqref{eq:comple_wn}, which in view of Remark~\ref{rem:pos_wn} are assumed to satisfy $\real k_j>0$ and $\imag k_j\geq 0$. 

The Helmholtz single-layer ($S_{\zeta,j}$), double-layer ($K_{\zeta,j}$), adjoint double-layer ($K'_{\zeta,j}$) and hypersingular ($N_{\zeta,j}$) operators are in turn defined as
\begin{equation}\begin{split} \lf(S_{\zeta,j}\varphi\rg)(\nex) := \int_{\Gamma} G_{\zeta,j}(\nex,\ney)\varphi(\ney)\de s(\ney), &\qquad \lf({K'}_{\zeta,j}\varphi\rg)(\nex) := \int_{\Gamma} \frac{\p G_{\zeta,j}(\nex,\ney)}{\p n(\nex)}\varphi(\ney)\de s(\ney),\\
\lf(K_{\zeta,j}\varphi\rg)(\nex) := \int_{\Gamma} \frac{\p G_{\zeta,j}(\nex,\ney)}{\p n(\ney)}\varphi(\ney)\de s(\ney),&\qquad
\lf(N_{\zeta,j}\varphi\rg)(\nex) := \mathrm{f.p.}\int_{\Gamma} \frac{\p^2 G_{\zeta,j}(\nex,\ney)}{\p n(\nex)\p n(\ney)}\varphi(\ney)\de s(\ney),\end{split}\label{eq:int_op}
\end{equation}
for  $\nex\in \Gamma$.  As usual, the initials f.p. in the definition of the hypersingular operator $N$ stand for Hadamard finite-part integral.  Throughout this section we assume that the layer potentials and boundary integral operators are defined for density functions $\varphi:\Gamma\to\C$ that make the  integrals in~\eqref{eq:lay_pot} and~\eqref{eq:int_op} conditionally convergent---after the regularizations needed due to the kernel singularities.

Unlike the seminal reference~\cite{bruno2016windowed} on the WGF method, where a direct integral formulation approach is followed, we use here an indirect formulation whose derivation is conceptually simpler.  Introducing two unknown density functions $\varphi_\zeta,\psi_\zeta:\Gamma\to\C,$ we  seek reflected and transmitted fields of the form 
\begin{equation}
u^{(j)}(\nex,\zeta) = \nu_j^{-1}(\mathcal D_{\zeta,j}\varphi_\zeta)(\nex)-(\mathcal S_{\zeta,j}\psi_\zeta)(\nex),\quad\nex\in\Omega_j,\quad j=1,2.\label{eq:layer_rep}
\end{equation}
Clearly, the potentials~\eqref{eq:layer_rep} satisfy the Helmholtz equation~\eqref{eq:helmholtz} in~$\Omega_j$. Enforcing the transmission conditions~\eqref{eq:trans_cond_1}-\eqref{eq:trans_cond_2} we readily arrive at the  integral equation system 
\begin{equation}\label{eq:int_eq_full}
-E\bol\phi_\zeta + T_\zeta[\bol\phi_\zeta] =\bol\phi^{\mathrm{inc}}_\zeta \quad \mbox{on}\quad \Gamma
\end{equation}
for the  vector density $\bol\phi_\zeta=[\varphi_\zeta,\psi_\zeta]^T$, where $ E = \frac{1}{2}\begin{bmatrix}
    \nu^{-1}_1+\nu^{-1}_2& 0\\
    0& \nu_1+\nu_2
	\end{bmatrix},$ $\bol\phi^{\mathrm{inc}}_\zeta = \begin{bmatrix}
f(\cdot,\zeta)\\
g(\cdot,\zeta)
	\end{bmatrix},
$ and
\begin{equation}
T_\zeta = \begin{bmatrix}
	\nu_2^{-1}K_{\zeta,2}-\nu_1^{-1}K_{\zeta,1}& -S_{\zeta,2}+ S_{\zeta,1}\smallskip\\
	N_{\zeta,2}-N_{\zeta,1}& - \nu_2{K'}_{\zeta,2}+ \nu_1{K'}_{\zeta,1}
	\end{bmatrix}.\label{eq:transmission_operator}
\end{equation}

Instead of solving~\eqref{eq:int_eq_full} on the entire unbounded material interface $\Gamma$, a locally
windowed problem is used to obtain the surface density functions $\varphi_\zeta$ and $\psi_\zeta$ over relevant portions of~$\Gamma$. In order to do so we
introduce a slow-rise smooth window function $w_A$ which is non-zero in an interval
of length~$2A$. In detail, our window function is given by  $w_A(x) = \eta(x,cA,A)$    where
\begin{equation}\label{eq:window_function}
\eta(t,t_0,t_1):= \lf\{\begin{array}{cll}1&\mbox{if}&|t|<t_0,\\
\displaystyle\exp\lf(\frac{2\e^{-1/u}}{u-1}\rg), u=\frac{|t|-t_0}{t_1-t_0},&\mbox{if}&t_0<|t|<t_1,\\
0&\mbox{if}&|t|\geq t_1,\end{array}\rg.
\end{equation}
where $0<c<1$. The width~$2A>0$ of the support of~$w_A$ is selected in such a way that $1-
w_A(x)$ vanishes on any corrugations that exist on the surface $\Gamma$.  Letting $W_{\!\!A}=w_A\cdot I$, where
$I$ is the $2\times 2$ identity matrix, we then  consider windowed integral  equation
\begin{equation}
-E \bol\phi_{\zeta,A}+  T_\zeta[W_{\!\!A} \bol\phi_{\zeta,A}] = \bol\phi^{\mathrm{inc}}_\zeta \quad\mbox{on}\quad \widetilde\Gamma_A=\{\nex\in\Gamma:w_A(x)\neq0\}.\label{eq:windowed_version}
\end{equation}
As is shown in~\cite{bruno2016windowed,perez2017windowed} the errors in the approximation $\bol\phi_\zeta\approx \bol\phi_{\zeta,A}$ in~$\Gamma_A=\{\nex\in\Gamma:w_A(x)=1\}$, for a fixed $\zeta$, decay  super-algebraically fast as the window size $A\to\infty$.  (Note that a single window size $A>0$ is here used for all the (eventually discrete) values of the variable $\zeta$.)

For a sufficiently smooth curve~$\Gamma,$ it can be easily shown that the windowed BIE~\eqref{eq:windowed_version} is of the second-kind for all $A>0$~\cite[Appendix D]{perez2017windowed}---as each one of the integral operators in the blocks of~$T_\zeta$ is given in terms of weakly singular kernels. Moreover, the windowed BIE~\eqref{eq:windowed_version} can be solved by means of any standard  BIE solver. 

With the approximate density functions $\varphi_{\zeta,A}$ and $\psi_{\zeta,A}$ at hand, the approximate reflected and transmitted fields  (in the frequency-domain) can be easily obtained by respectively substituting $\varphi_\zeta$ and $\psi_\zeta$ by $\varphi_{\zeta,A}$ and $\psi_{\zeta,A}$ in the representation formula~\eqref{eq:layer_rep}. These substitutions produce the approximate fields 
 \begin{equation} \label{eq:window_RF}
u^{(j)}_{A}(\nex,\zeta)= \nu_j^{-1}\mathcal D_{\zeta,j}\left[w_A\varphi_{\zeta,A}\right](\nex)-\mathcal S_{\zeta,j}\left[w_A\psi_{\zeta,A}\right](\nex),\quad\nex\in\Omega_j,\quad j=1,2,
\end{equation}
which, for each fixed $\zeta$, exhibit errors that decay super-algebraically as $A\to\infty$ in the regions $\Omega_{j,A}=\{\nex=(x,y)\in\Omega_j:w_A(x)=1\}$.

\begin{remark} Note that the use of a single window size $A$ allows for the numerous windowed BIEs~\eqref{eq:windowed_version} (which, in view of Remark~\ref{rem:pos_wn}, are $\sim\!\!N/2$ in total) to be solved using a single discretization of the curve $\widetilde \Gamma_A$, which can also be used in the numerical evaluation of  the windowed representation formulae~\eqref{eq:window_RF}. 
\end{remark}

\section{Examples and applications}\label{sec:num} 
This section presents a variety of numerical examples that demonstrate the accuracy of the proposed convolution quadrature method for problems of scattering in the presence of the unbounded material interfaces.

\subsection{Nystr\"om method and frequency-domain problems}
Throughout this paper we utilize a high-order Nystr\"om method for the spatial discretization of the  windowed frequency-domain BIEs~\eqref{eq:windowed_version}. Nystr\"om methods enjoy well-known advantages over other boundary integral equation methods. Unlike BEMs, for instance, Nystr\"om methods require numerical evaluation of only one boundary integral integral per grid point. Furthermore, they can easily yield high-order convergence without compromising the computational cost. Those advantages become even more apparent in CQ calculations where large numbers of frequency-domain problems are typically needed to be solved all at once.

Among the many two-dimensional  Nystr\"om methods available in the literature, we use here the one based on the Alpert quadrature rule~\cite{alpert1999hybrid} of order sixteen. This quadrature rule is designed to deal with weak singularities such as the logarithmic singularities present in the integral kernels defining the operator $T_\zeta$ in~\eqref{eq:transmission_operator}.  This BIE method enjoys two immediate advantages over, say,  the classical spectrally-accurate Martensen-Kussmaul (MK) Nystr\"om method~\cite[section 3.5]{colton2012inverse} (for the kind of problems  concerning this work), which is arguably the best discretization method available for frequency-domain problems. On one hand, the Alpert-based Nystr\"om method can easily handle large complex wavenumbers, such as those produced by CQ methods, and, on the other hand, it is compatible with the Fast Multipole Method~\cite{Hao:2013do}. (As is well-known, for complex wavenumbers, the MK method suffers from numerical instabilities arising from round-off errors~\cite{lu2014efficient,wang1998modal}.)

 \begin{figure}[h!]
\centering	
 \subfloat[][Convergence of the Alpert-based Nystr\"om  solver.]{\includegraphics[scale=0.65]{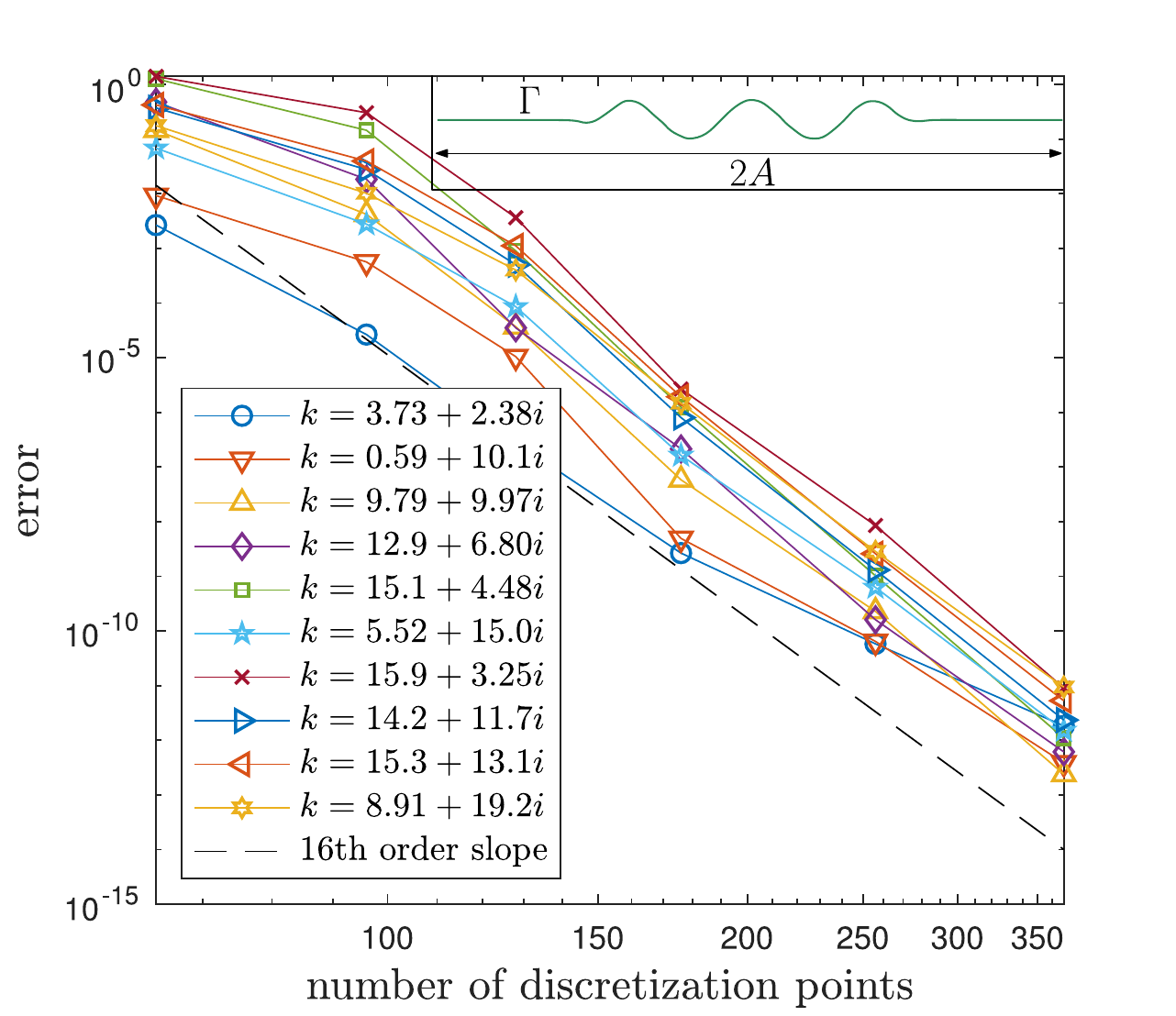}\label{fig:conv_N_a}}
 \subfloat[][Convergence of the WGF method.]{\includegraphics[scale=0.65]{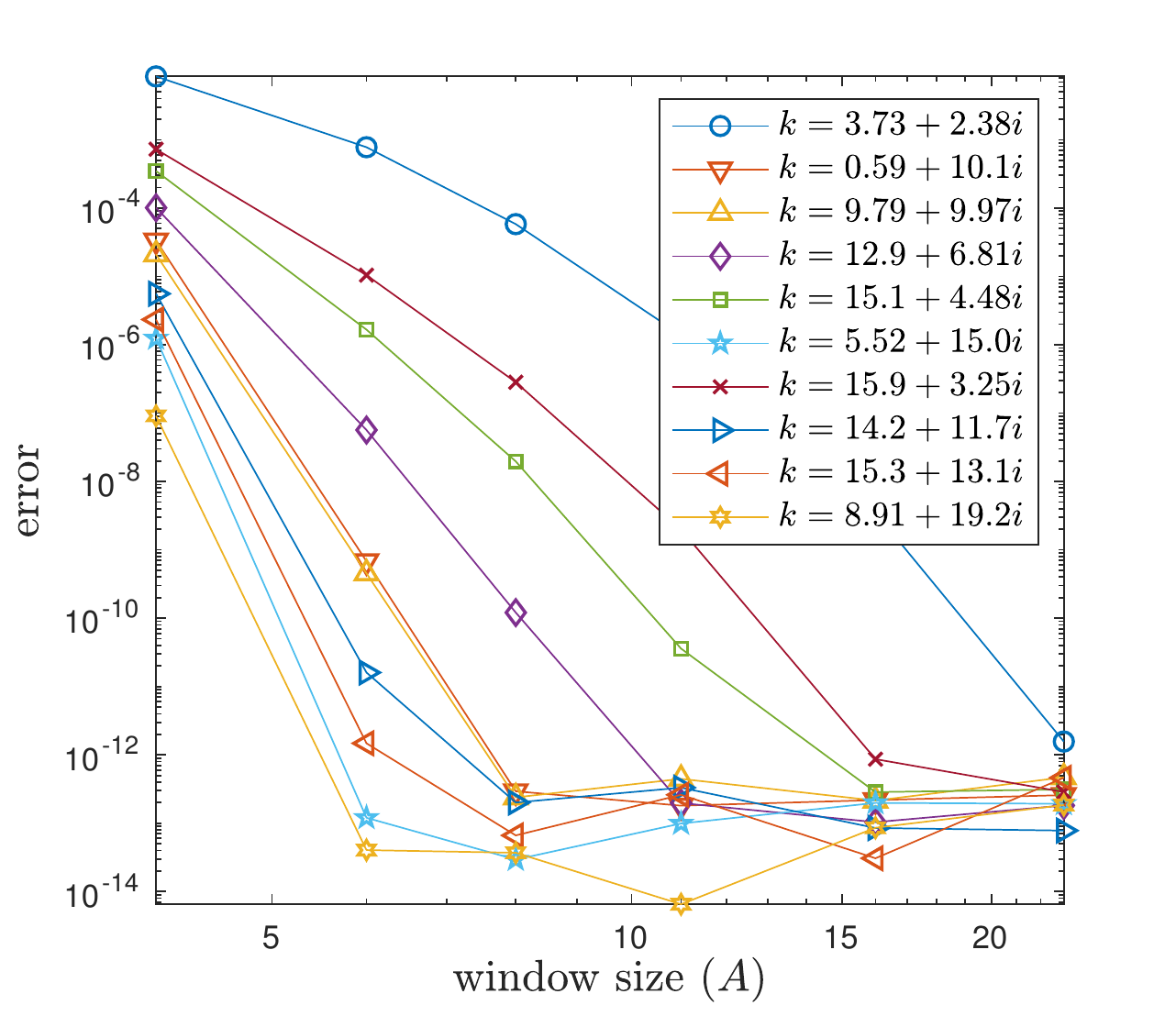}\label{fig:conv_A_b}}
\caption{Errors in the solution of the problem ten frequency-domain problems with complex wavenumbers $k_1=k$ and $k_2=k/2$ with $k$ selected randomly within the domain  $\{0<\real z,\imag z<20\}$ obtained  by (a) varying the number of discretization points for fixed window size, and (b) varying the window size $A$ for a fix number of discretization points. The curve $\Gamma$ is shown in the inlet figure.The right-hand-side $[f,g]^T$ in the windowed integral equation was selected as $f= v$ and $g=\p_n v$ on $\Gamma$ where $v(x,y)=\e^{-ik_1y}$.}\label{fig:convergenve_FD}
\end{figure}  

In order to validate our frequency-domain Nystr\"om solver for the solution of the windowed BIE~\eqref{eq:windowed_version}, we consider a two-layer medium with a smooth cosine-shaped defect. In detail, the penetrable interface considered is  
\begin{equation}\label{eq:cos_interface}
\Gamma=\lf\{\lf(t,\frac12\cos(2t)\eta(t,2.5,5)\rg)\in\R^2,t\in\R\rg\},
\end{equation} 
where $\eta$ is defined in~\eqref{eq:window_function} (the curve $\Gamma$ is depicted in the inset in Figure~\ref{fig:conv_N_a}). The wavenumbers considered are $k_1=k$ and $k_2=k/2$ in $\Omega_1$ and $\Omega_2$, respectively, for ten different values of the parameters $k$ which were selected randomly from the set $\{0<\real z,\imag z<20\}$ by a uniform distribution. These wavenumbers are meant to be representative of those generated by CQ methods.  The windowed BIE~\eqref{eq:windowed_version} is then numerically solved for each one of the randomly selected $k$ values for various numbers of discretization points and for a fixed window size $A=8$. The numerical errors in the fields are displayed in Figure~\ref{fig:conv_N_a} where it can be clearly seen that our solver yields the expected order of convergence.  The numerical error is here defined as $\max\{e^{(1)},e^{(2)}\}$ where 
\begin{equation}\label{eq:error_formula}
e^{(j)}(\cdot)=\max_{i=1,2,3}\lf|u^{(j)}_{A}(\nex^{(j)}_i,\cdot)-\tilde u^{(j)}_{A}(\nex^{(j)}_i,\cdot)\rg|/\max_{i=1,2,3}\lf|\tilde u_{A}(\nex^{(j)}_i,\cdot)\rg|,\quad j=1,2,
\end{equation}
and where the evaluation points are $\nex^{(1)}_1=(-1,1)$, $\nex^{(1)}_2=(0,1)$ and $\nex^{(1)}_3=(1,1)$ in the upper domain $\Omega_1$, and  $\nex^{(2)}_1=(-1,-1)$, $\nex^{(2)}_2=(0,-1)$ and  $\nex^{(2)}_3=(1,-1)$ in the lower domain $\Omega_2$. Both the sample fields $u^{(j)}_A$ and the reference fields $\tilde u^{(j)}_A$ in~\eqref{eq:error_formula} were obtained by numerically solving the windowed BIE~\eqref{eq:windowed_version} and then evaluating the representation formula~\eqref{eq:window_RF}, using the approximate surface densities. The reference fields ($\tilde u^{(j)}$) were produced using a fine grid consisting of~$512$ discretization points.

In order to demonstrate the high-order convergence of the WGF method in the context of CQ methods, we next consider the frequency-domain problems of the previous example but they are now solved for various window sizes $A>0$. The number of discretization points used in this example is such that $\sim\!\!15$ per unit length are used, which turns out to be enough to guarantee that the dominant error in all the calculations stems from the use of a finite window size $A>0$ and not from the Nystr\"om discretization of the windowed BIE~\eqref{eq:windowed_version}. 
Figure~\ref{fig:conv_A_b} displays the numerical errors obtained for the various window sizes and complex wavenumbers considered. The error is measured as in the previous example but with the reference fields produced using a large window size  $A = 32$. As expected, super-algebraic convergence is observed for all the complex wavenumbers considered, with error curves exhibiting a strong dependence on the wavenumber; faster convergence is observed for wavenumbers with larger imaginary part (this is partly explained by the fast (exponential) decay of the integral kernels). On the other hand, for a fixed imaginary part, faster convergence is expected for wavenumbers with larger real part~\cite{bruno2016windowed,perez2017windowed}.

The fact that the convergence of the WGF method depends on the wavenumber rises the issue of selecting a single appropriate window size to be used in the solution of all the frequency-domain problems. This issue can be easily resolved in the case of BDF-based CQ methods by noticing that the  wavenumber with the smallest imaginary part is also the wavenumber with the smallest real part.  This is due to the fact that the CQ-produced wavenumbers~\eqref{eq:comple_wn} lie on the boundary of a bounded convex set contained in the upper complex half-plane. Therefore, in order to achieve aceptable WGF errors in the solution of all the frequency-domain problems, it suffices to select $A>0$ large enough so that the WGF errors in the solution of the problem with smallest wavenumber is acceptable.  This procedure is utilized in the selection of the window-size parameter $A$ in all the time-domain problems considered in this paper.  

\begin{remark}
An alternative approach to deal with the frequency-dependent convergence of the WGF method can be devised for simple problems for which discretizations of the interfaces can be inexpensively produced. Since the actual CQ-WGF approximation to wave-equation solution at a point $\nex$ is a linear combinations of the fields~\eqref{eq:window_RF} resulting from a discrete set of $\zeta$ values, a $\zeta$-dependent windowed sizes $A_\zeta$ can in principle be used in the numerical solution of the windowed BIE and in the evaluation of the windowed representation formulae~\eqref{eq:window_RF}. This procedure would allow to eliminate  inefficiencies stemming from both the use of unnecessarily large values of $A$ for large wavenumbers, and from the use of over discretized spatial grids for small wavenumbers. 
\end{remark}

\subsection{Time-domain scattering problems}\label{sec:CQ_results}
This section encompasses several challenging examples that validate our CQ-WGF method for the solution of time-domain scattering problems. For the sake of definiteness, in what follows we consider the CQ method associated to backward differences of orders two (BDF2) with corresponding polynomial $\gamma(\zeta)= \frac{1}{2}(\zeta^2-4\zeta+3)$ (higher-order CQ methods can be easily incorporated). Following \cite{Banjai:2009in} the radius of the circular contour in~\eqref{eq:inv_Z} is selected as $\lambda=\epsilon^{\frac{1}{2N}}$ in all convolution quadrature computations, where $ \epsilon>0$ denotes the machine-precission number and  where $N$  is the total number of time-steps. 
\begin{figure}[h!]
\centering 
\subfloat[][]{\includegraphics[scale=0.65]{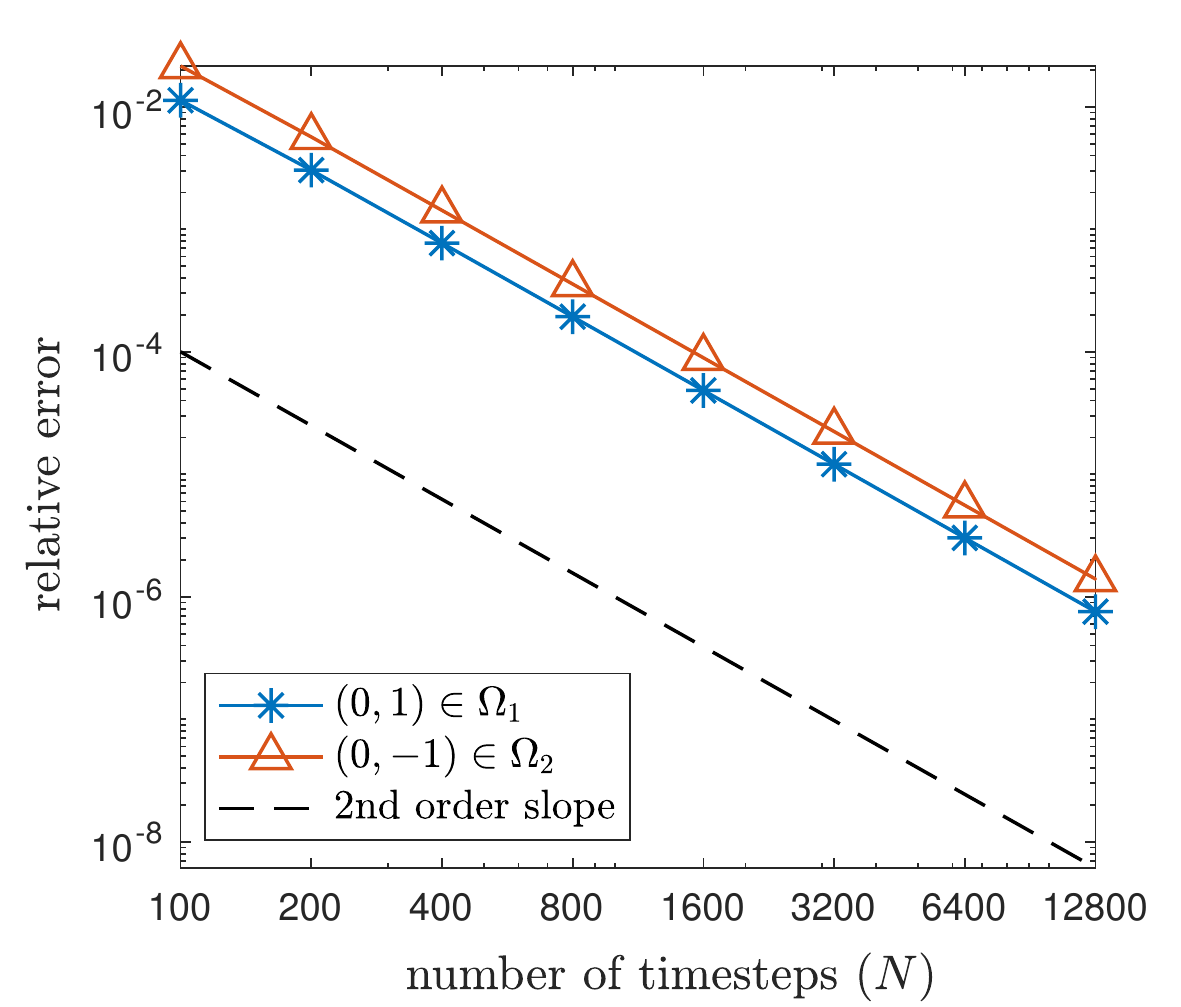}\label{fig:ex2_a}} 
\subfloat[][]{\includegraphics[scale=0.75]{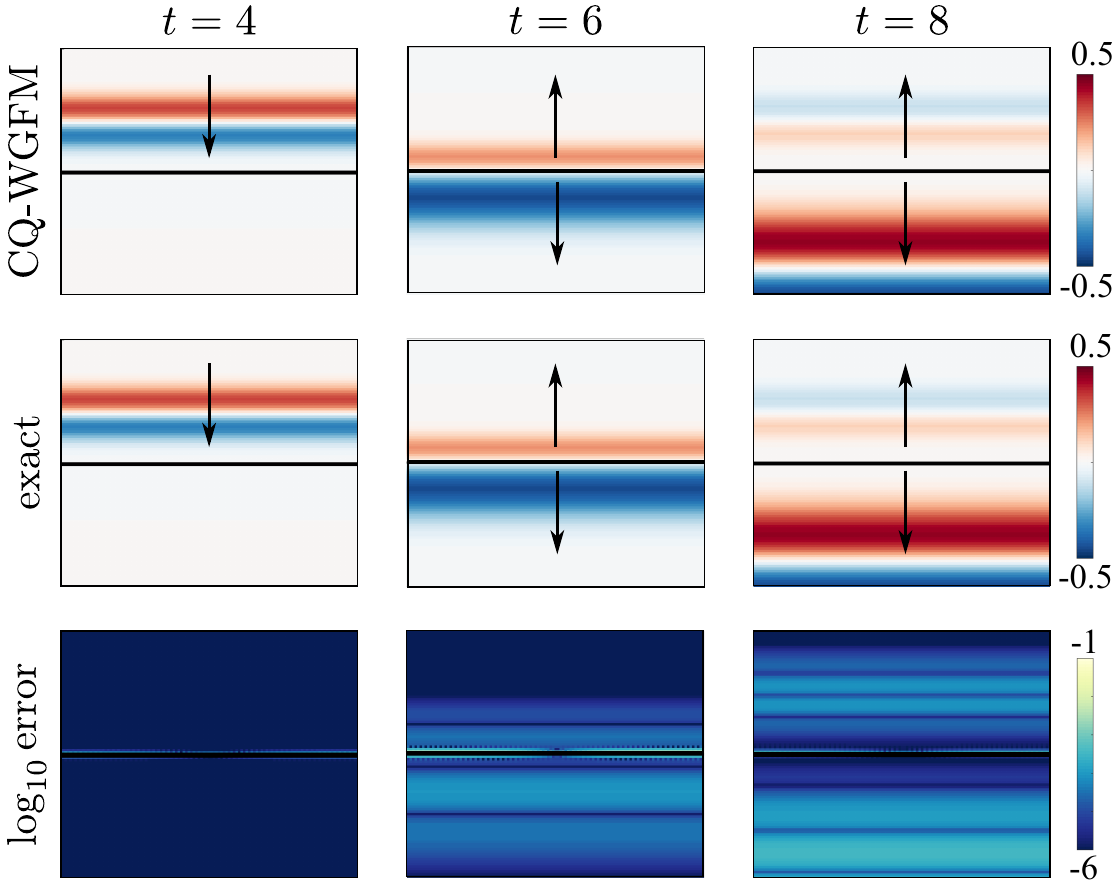}\label{fig:ex2_b}}
\caption{Comparison of the CQ-WGF solution with an exact solution. (a) Relative errors---measured at the final time $ T = 10 $  at the observation points $(0, 1)\in\Omega_1$ and $(0, -1)\in\Omega_2$---in the approximate wave-equation solution obtained using the sixteenth-order Alpert-based Nystr\"om method and the CQ-BDF2 method. (b) Three snapshots of the approximate (first row) and exact (second row) total fields as well as log-10 of the absolute value of their difference (third row), where each one of the fields is plotted within the domain~$[-6,6]\times [-5,5]\subset\R^2$. The black arrows in indicate the direction of the incoming incident planewave (first column) as well as the transmitted and reflected planewaves (second and third columns) generated at the planar penetrable interface.} \label{fig:tdplanewave}
\end{figure}

\paragraph{Planar two-layer medium.} In the first example of this section we consider the scattering of a planewave off of a planar two-layer medium consisting of the subdomains $\Omega_1=\R^2_+$ and $\Omega_2=\R^2_-$ with wavenumbers $c_1$ and $c_2$, respectively, for which the exact solution can be analytically constructed from Snell's law and Fresnel equations~\cite{brekhovskikh2013acoustics}. In detail, for a general incident planewave of the form $U^\inc(\nex,t)= f(c_1 (t - t_{\text{lag}}) - \nex\cdot \bol{d}(\theta^\inc)),$ with $\bol d(\theta)=(\cos\theta,-\sin\theta)$ and $\theta\in [0,\pi]$, the exact total field solution of the problem of scattering is given by 
$$
U(\nex,t) = \lf\{\begin{array}{ccc}U^\inc(\nex,t)+R(\theta^\inc)f\lf(c_1 (t - t_{\text{lag}}) - \nex\cdot \bol{d}(-\theta^\inc)\rg),& \nex\in\Omega_1=\R^2_+,\smallskip\\
T(\theta^\inc)f\lf(c_2 (t - t_{\text{lag}}) - \nex\cdot \bol{d}(\theta^{\rm ref})\rg),&\nex\in\Omega_2=\R^2_{-},\end{array}\rg.
$$
where the refraction  angle $\theta^{\rm ref}\in[0,\pi]$ (measured with respect to the horizontal) is determined by the relation $n:=c_1/c_2= \cos(\theta^\inc)/\cos(\theta^{\rm ref})$ and where the reflection ($R$) and transmission $(T)$ coefficients are given by 
 \begin{equation}
R(\theta^\inc) = \dfrac{\sin \theta^\inc - \sqrt{n^2 - \cos^2 \theta^\inc}}{\sin\theta^\inc + \sqrt{n^2 - \cos^2 \theta^\inc}} \andtext  T(\theta^\inc)=1 + R(\theta^\inc).
 \end{equation}
In this particular example we consider  $f(t) = \sin(t)  \exp(-\sigma t^2) $, $\theta^\inc=\pi/2$, $\sigma = 1.5$, $t_{\text{lag}} = 5$, and the wavespeeds $c_1 = 1$ and $c_2 = 2$. 
The numerical errors produced  by proposed CQ-WGF procedure are displayed in Figure~\ref{fig:ex2_a} where the expected second-order convergence in time of the fields in each of the layers can be observed. A window size of $A=40$ and a total number of $400$ discretization points were used in the numerical solution of each of the windowed BIEs~\eqref{eq:windowed_version}. These parameters were selected so as to guarantee spatial errors below $10^{-6}$ at the observation points considered, in all the frequency-domain solutions for the complex wavenumbers produced by the CQ-BDF2 method. The approximate, exact and the logarithm in base ten of the absolute value of their difference are displayed in Figure~\ref{fig:ex2_b}.

\begin{figure}[h!]
	\centering 
\subfloat[][]{\includegraphics[scale=0.65]{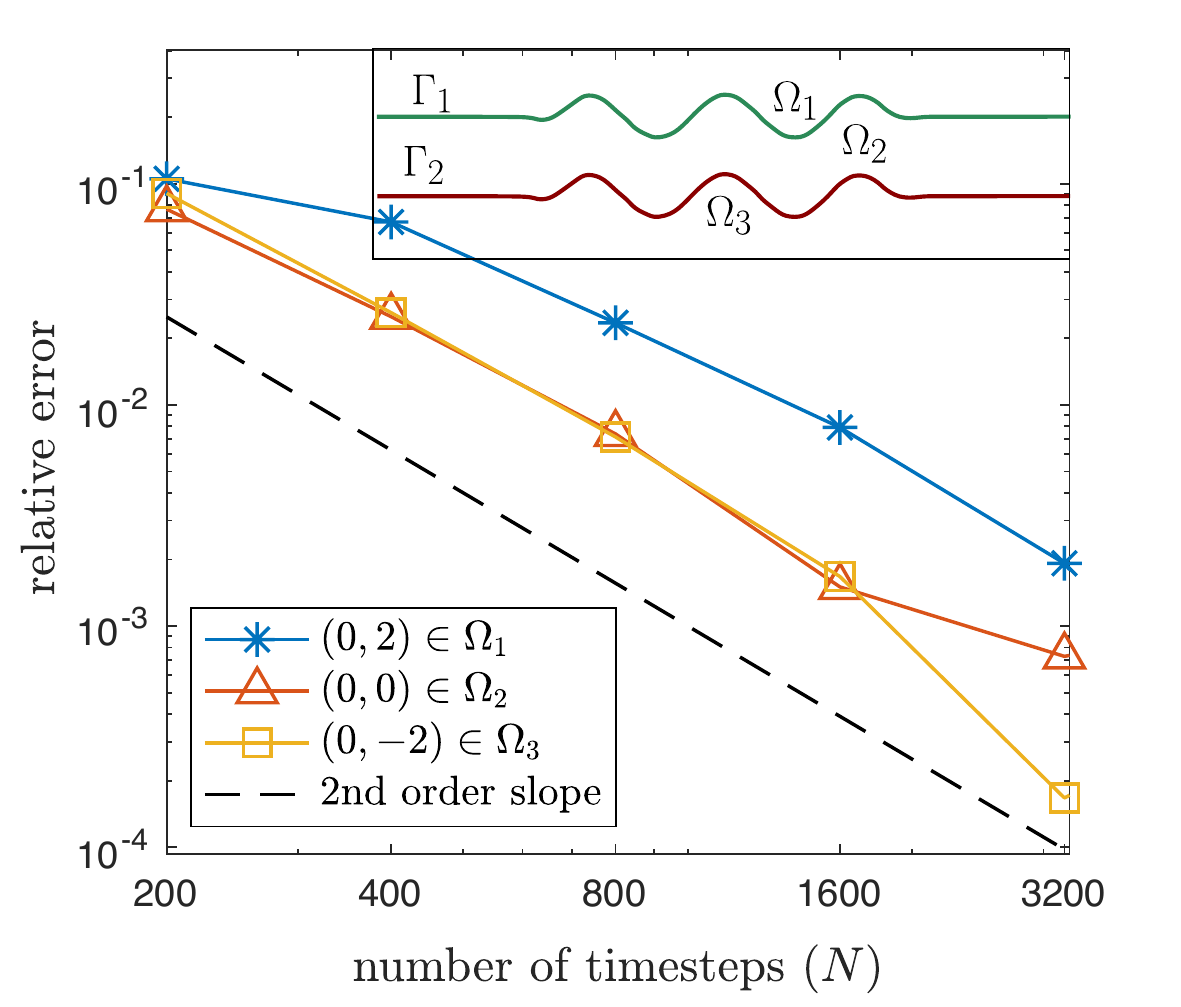}\label{fig:3layer_a}}
\subfloat[][]{\includegraphics[scale=0.9]{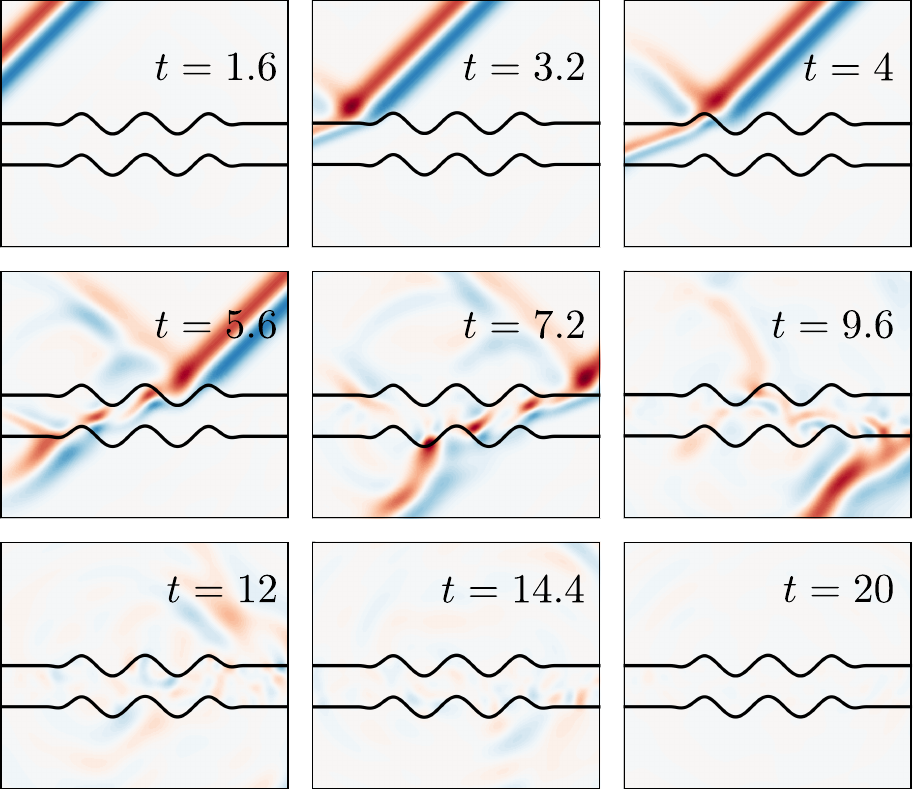}\label{fig:3layer_b}}
	\caption{Scattering of a planewave off of a three-layer medium. (a) Relative errors---measured at the final time $ T = 20 $  at the observation points $(0, 2)\in\Omega_1$, $(0, 0)\in\Omega_2$ and $(0,-2)\in\Omega_3$---in the approximate CQ-WGF  wave-equation  solution obtained using the sixteenth-order Alpert-based Nystr\"om method and the CQ-BDF2 method. (b) Nine snapshots of the solution.} \label{fig:tdplanewave_2}
\end{figure}

\paragraph{Multi-layer medium.}
In the second example of this section we consider a three-layer medium with penetrable interfaces $\Gamma_1$ and $\Gamma_2$ defined as  $\Gamma_1=\Gamma$ and $\Gamma_2=\Gamma+\{(0,-2)\}$, where $\Gamma$ is the curve defined in~\eqref{eq:cos_interface} and depicted in the inset in Figure~\ref{fig:conv_N_a}. The wavespeeds are $c_1 = 2$, $c_2 = 1$ and $c_3 = 2$ in $\Omega_1$, $\Omega_2$ and $\Omega_3$, respectively (the various domains and interfaces involved in this problem are displayed in the inset of Figure~\ref{fig:3layer_a}). As in the previous example, the incident field is the planewave $U^\inc(\nex,t)=f(c_1 (t - t_{\text{lag}}) - \nex\cdot \bol{d}(\theta^\inc))$ with parameters $ \theta^\inc = \frac\pi4$, $\sigma = 1.5$ and $t_{\rm{lag}} = 5$. Convergence results are shown in Figure~\ref{fig:3layer_a} and snapshots of the solution are displayed in Figure~\ref{fig:3layer_b}. The derivation of the corresponding BIE in this case is completely analogous to the one presented in Section~\ref{sec:WGFM} above for the two-layer problem. The widow size $A=15$ and a total of $240$ discretization points on each interface were used in the WGF solution of the frequency-domain problems. The reference fields were obtained using $N=6400$ timesteps.

\paragraph{Waveguide and waveguide branches.}
Finally, we consider two different waveguide problems. The incident field is selected as a causal periodic pulse placed at a point $\nex_0$ within the waveguide structure---which in both cases is denoted by $\Omega_2$. In detail, the incident field is given by the time convolution
\begin{equation}\label{eq:sourcepulse}
U^\inc(\nex, t) := \int_{0}^{t}   G(\nex,\nex_0, t-\tau) f(\tau) \ \text{d}\tau,\qquad t\geq0,\quad (\nex_0\in\Omega_2)
\end{equation}
of the fundamental solution of the wave equation~\cite{sayas2016retarded} 
$$ G(\nex, \ney, t) = \dfrac{H\lf(t-c_2^{-1}|\nex - \ney|\rg)}{2\pi \sqrt{t^2 - c_2^{-2}|\nex - \ney|^2}},$$ 
where $H$ denotes the Heaviside step function, and the periodic signal $f(t) = \sin(2 t)$. The convolution integral~\eqref{eq:sourcepulse} is evaluated by means of BDF2-based CQ method~\cite{hassell2016convolution}.

\begin{figure}[h!]
	\centering 
\subfloat[][]{\includegraphics[scale=0.65]{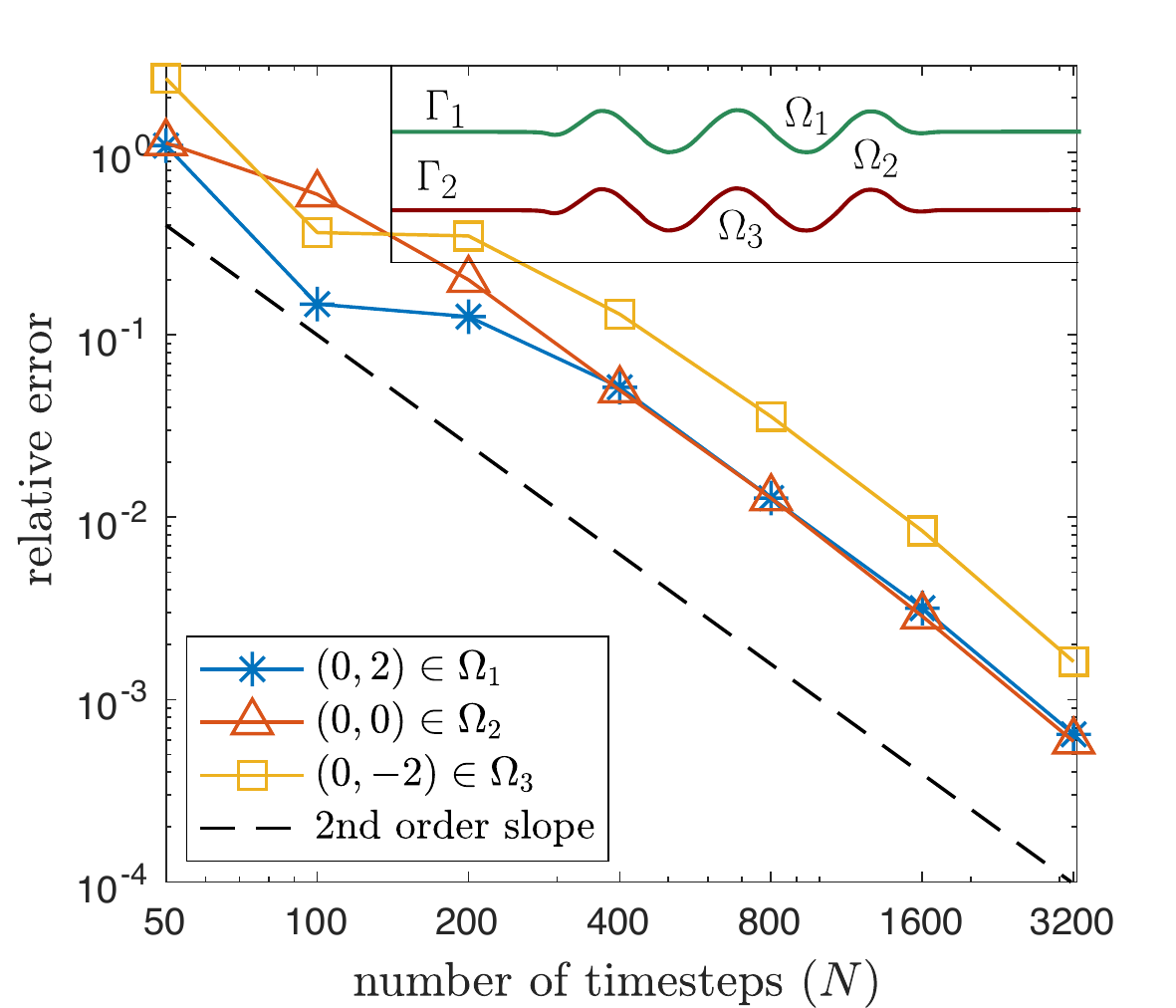}\label{fig:WG_a}}
 \subfloat[][]{\includegraphics[scale=0.9]{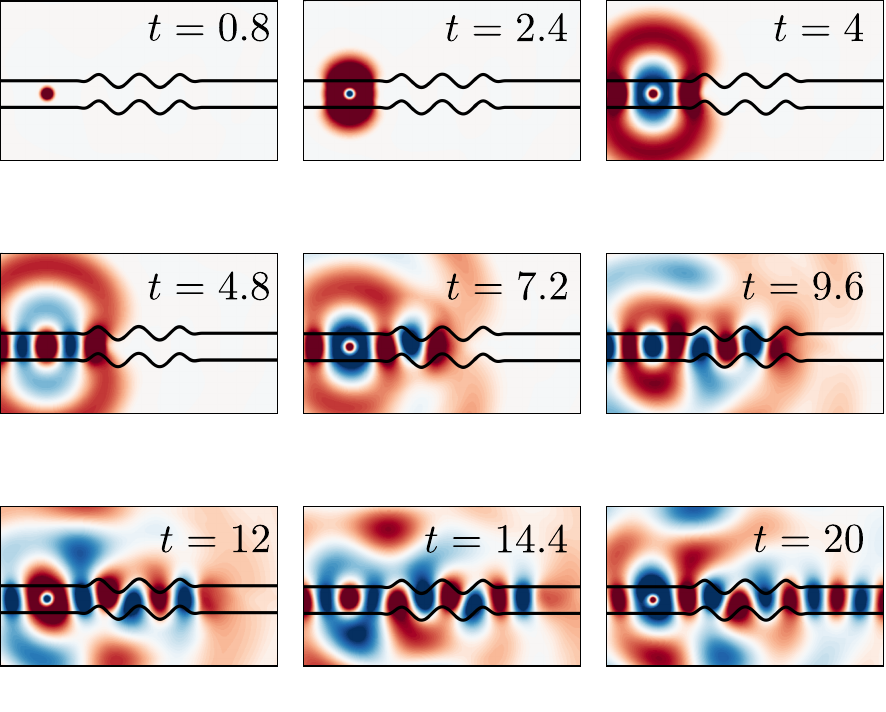}\label{fig:WG_b}}
	\caption{Wave propagation within an open waveguide. (a) Relative errors---measured at the final time $ T = 20 $  at the observation points $(0, 2)\in\Omega_1$, $(0, 0)\in\Omega_2$ and $(0,-2)\in\Omega_3$---in the approximate CQ-WGF  wave-equation  solution obtained using the sixteenth-order Alpert-based Nystr\"om method and the CQ-BDF2 method. (b) Nine snapshots of the solution.} \label{fig:tdwaveguide}
\end{figure}

The geometry of our first waveguide problem---which is the same considered in the previous example---is depicted in the inset of Figure~\ref{fig:WG_a}. The wavespeeds are once again  $c_1 = 2, c_2 = 1$ and  $c_3 = 2 $ in $\Omega_1$, $\Omega_2$ and $\Omega_3$, respectively. The second-order convergence of the proposed methodology is demonstrated in Figure~\ref{fig:WG_a} where relative errors at three different points---one in each subdomain---are shown for various time discretizations. A fixed window size $A=15$ and a fixed number of discretization points (equal to 240) were used on each interface in the numerical solution of each of the corresponding windowed BIEs. The reference fields at the final time $T=20$, were obtained using $N=6400$ timesteps. Snapshots of the solution are displayed in Figure~\ref{fig:WG_b}. As expected, the time harmonic incident field considered eventually excites the first propagative mode of the waveguide that can be clearly seen in the last snapshot.

\begin{figure}[h!]
	\centering
\includegraphics[height=14cm]{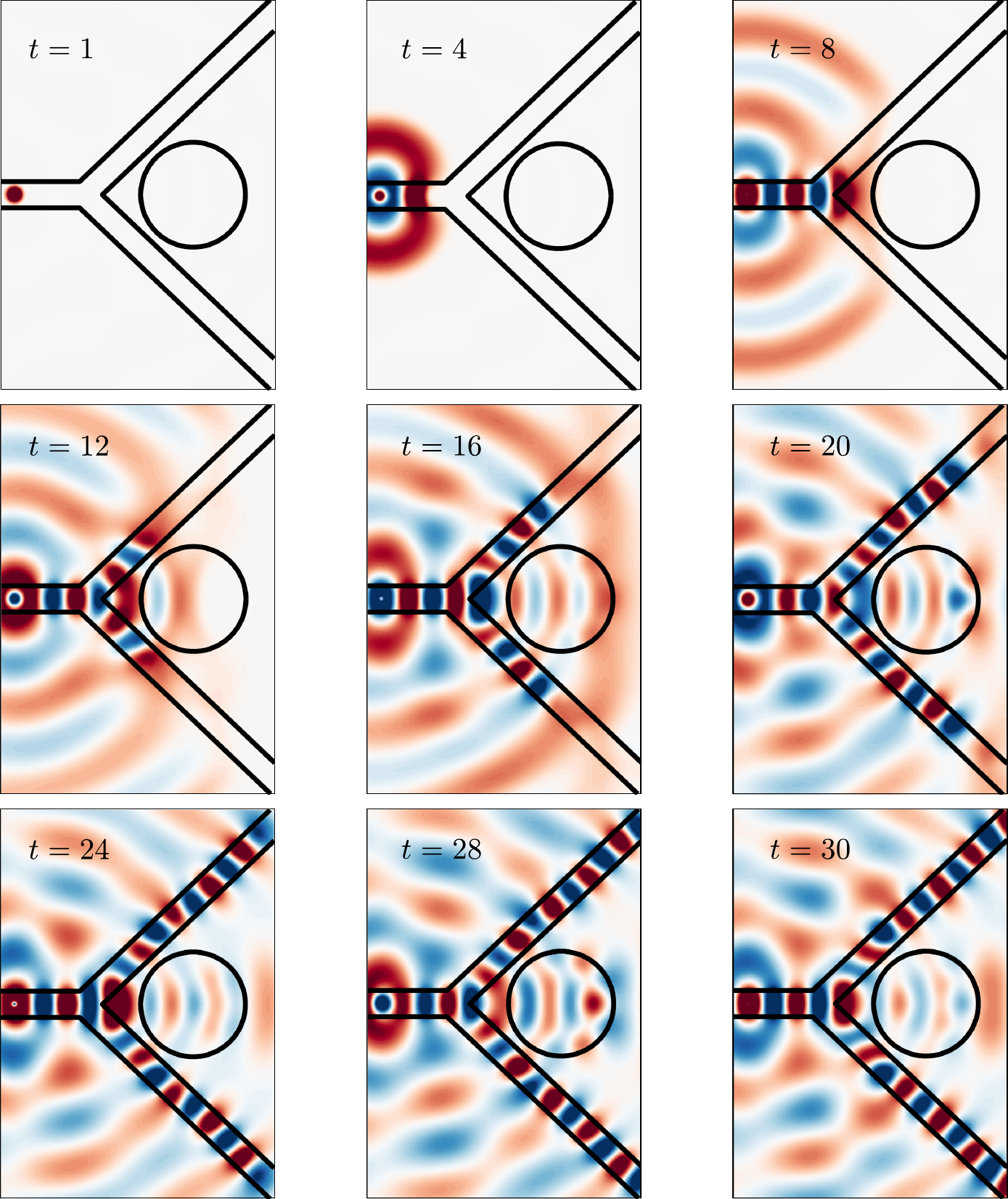}
	\caption{Wave propagation within a (non-smooth) waveguide branch.\label{fig:plotwaveguide2}} 
\end{figure}

In our final example we consider a more complicated example consisting of a waveguide branch and a circular resonator. Note that  some of the interfaces are not smooth. In order to properly resolve the BIE densities and the fields near the corners, a sigmoid transformation is used to produce grids that accumulate discretization points near the corners thus ensure the overall high-order convergence of our frequency domain BIE solver~\cite{anand2012well,dominguez2016well}. The wavespeed in the waveguide and the resonator is $c =1$. Outside the waveguide and the resonator the wavespeed is $c=2$. Snapshots of the solution are presented in Figure \ref{fig:plotwaveguide2} which shows that a propagative mode is excited within the waveguide and it splits as it propagates into the two waveguide branches.

%
%
\appendix

\newpage
\bibliographystyle{abbrv}
\bibliography{References}
\end{document}